%% file: Main.tex
\documentclass[sn-mathphys,Numbered]{sn-jnl}

\usepackage{amsmath,amssymb,amsfonts}%
\usepackage{amsthm}%
\usepackage{mathrsfs}%
\usepackage[title]{appendix}%
\usepackage{textcomp}%
\usepackage{manyfoot}%
\usepackage{booktabs}%
\usepackage{algorithm}%
\usepackage{algorithmicx}%
\usepackage{algpseudocode}%
\usepackage{listings}%

\usepackage{booktabs} 
\usepackage{graphicx}
\usepackage{xspace}
\usepackage[normalem]{ulem}
\useunder{\uline}{\ul}{}
\usepackage{multirow}
\usepackage{lscape}
\usepackage{rotating}
\usepackage{supertabular}
\usepackage{longtable}
\usepackage{enumerate}
\usepackage{comment}
\usepackage{ragged2e}
\usepackage{array}
\PassOptionsToPackage{hyphens}{url}
\usepackage{url}
\usepackage{hyperref}
\usepackage{array}
\usepackage{ragged2e}
\usepackage{caption}
\newcolumntype{P}[1]{>{\RaggedRight\hspace{0pt}}p{#1}}
\usepackage{fontawesome}
\usepackage[table,xcdraw]{xcolor}

\usepackage{tcolorbox}
\usepackage[switch]{lineno}
\usepackage{xcolor,colortbl}

\newif\ifdraft
\drafttrue

\theoremstyle{thmstyleone}%
\theoremstyle{thmstyletwo}%

\theoremstyle{thmstylethree}%

\begin{document}
\input{Comments}

\title[Article Title]{AI for All: Identifying AI incidents Related to Diversity and Inclusion}

    
\author*[1]{\fnm{Rifat Ara} \sur{Shams}}\email{rifat.shams@csiro.au}

\author[1]{\fnm{Didar} \sur{Zowghi}}\email{didar.zowghi@csiro.au}

\author[1]{\fnm{Muneera} \sur{Bano}}\email{muneera.bano@csiro.au}

\affil[1]{\orgname{CSIRO's Data61}, \orgaddress{\country{Australia}}}


\abstract{The rapid expansion of Artificial Intelligence (AI) technologies has introduced both significant advancements and challenges, with diversity and inclusion (D\&I) emerging as a critical concern. Addressing D\&I in AI is essential to reduce biases and discrimination, enhance fairness, and prevent adverse societal impacts. Despite its importance, D\&I considerations are often overlooked, resulting in incidents marked by built-in biases and ethical dilemmas. Analyzing AI incidents through a D\&I lens is crucial for identifying causes of biases and developing strategies to mitigate them, ensuring fairer and more equitable AI technologies. However, systematic investigations of D\&I-related AI incidents are scarce. This study addresses these challenges by identifying and understanding D\&I issues within AI systems through a manual analysis of AI incident databases (AIID and AIAAIC). The research develops a decision tree to investigate D\&I issues tied to AI incidents and populate a public repository of D\&I-related AI incidents. The decision tree was validated through a card sorting exercise and focus group discussions. The research demonstrates that almost half of the analyzed AI incidents are related to D\&I, with a notable predominance of racial, gender, and age discrimination. The decision tree and resulting public repository aim to foster further research and responsible AI practices, promoting the development of inclusive and equitable AI systems.}




\maketitle

\input{Sections/1_Introduction}
\input{Sections/2_Background}
\input{Sections/3_Methodology}
\input{Sections/4_Results}
\input{Sections/5_Discussion}
\input{Sections/6_Threats_to_Validity}
\input{Sections/7_Conclusions}


\bibliography{References.bib}


\clearpage
\appendix
\section{Appendix A. Evolution of the Decision Tree}
\label{appendix_A}

    \begin{figure*}[!htbp]
            \centering
            \includegraphics[width=1\textwidth]{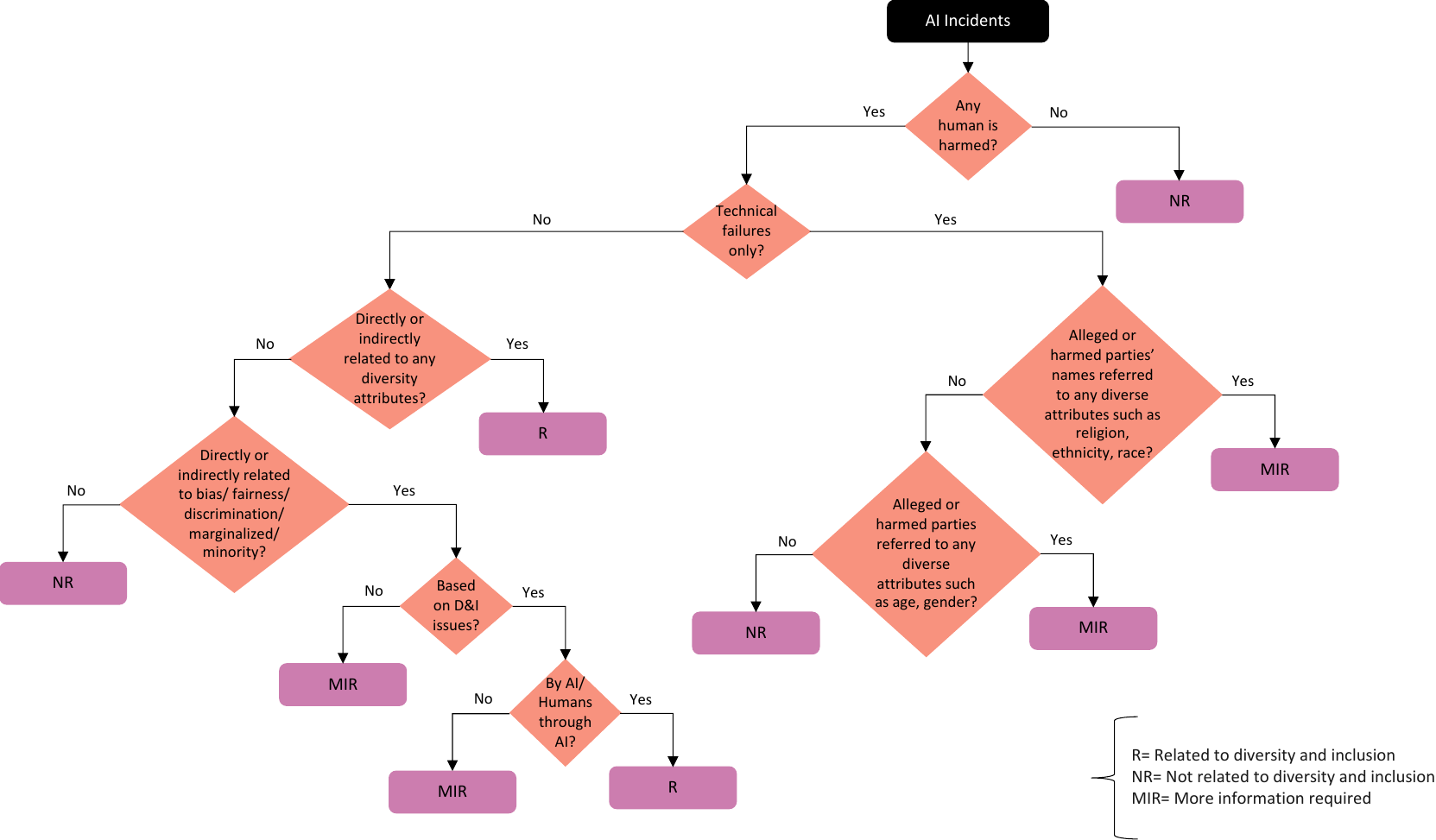}
            \caption{Decision tree: Version 1}
            \label{fig:decision_tree_v1}
    \end{figure*}

    \begin{figure*}[!htbp]
            \centering
            \includegraphics[width=1\textwidth]{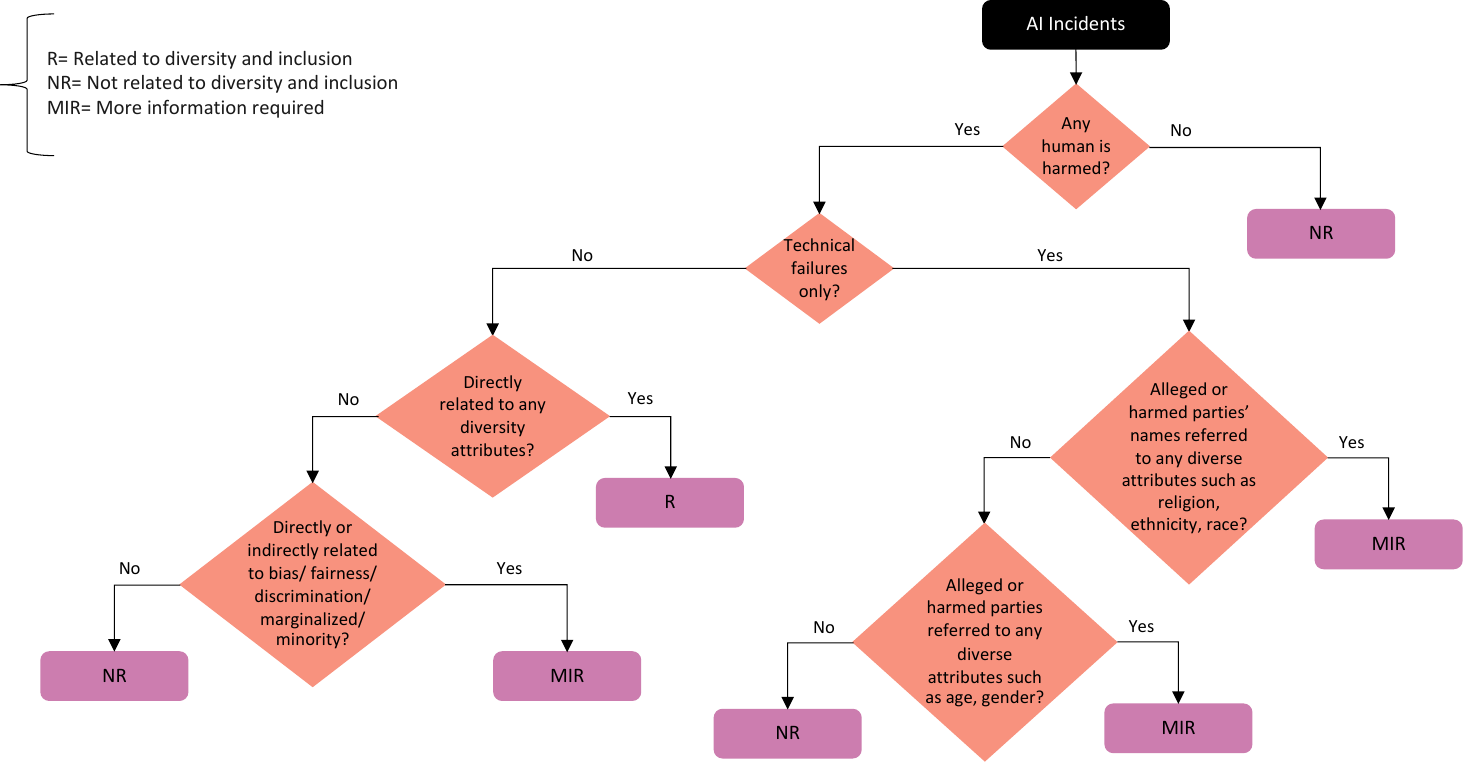}
            \caption{Decision tree: Version 2}
            \label{fig:decision_tree_v2}
    \end{figure*}

    \begin{figure*}[!htbp]
            \centering
            \includegraphics[width=0.85\textwidth]{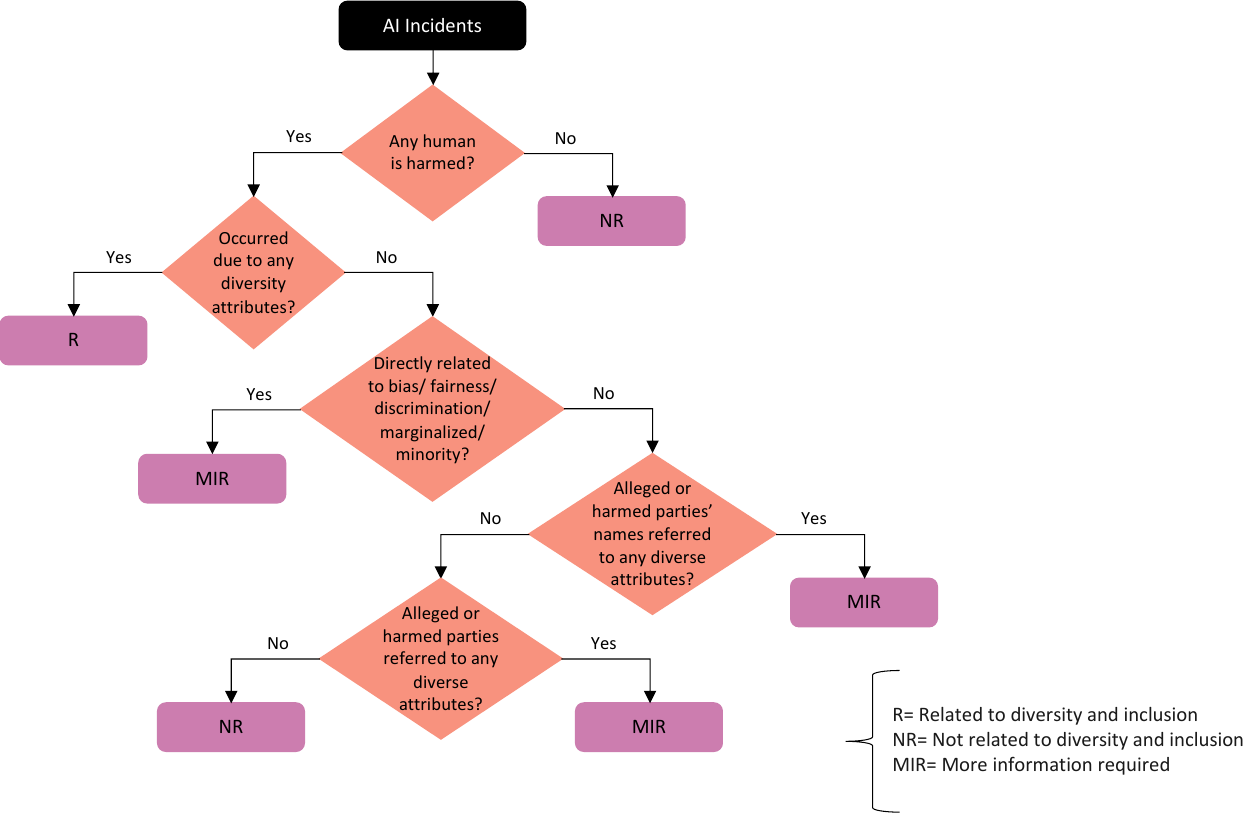}
            \caption{Decision tree: Version 3}
            \label{fig:decision_tree_v3}
    \end{figure*}

    \begin{figure*}[!htbp]
            \centering
            \includegraphics[width=0.7\textwidth]{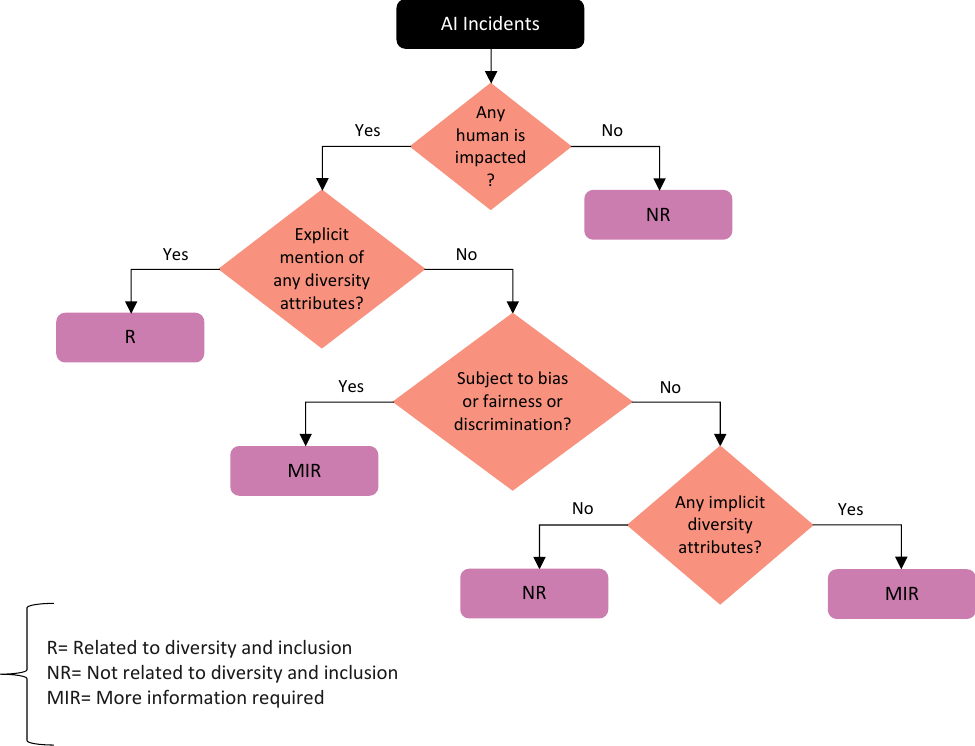}
            \caption{Decision tree: Version 4}
            \label{fig:decision_tree_v4}
    \end{figure*}
\clearpage

\end{document}

%% file: Comments.tex
\ifdraft
  \newcommand{\didar}[1]{{\color{red}\emph{Didar: #1}}\xspace}
  \newcommand{\muneera}[1]{{\color{blue}\emph{Muneera: #1}}\xspace}
  \newcommand{\fra}[1]{{\color{magenta}\emph{Fra: #1}}\xspace}
  \newcommand{\rifat}[1]{{\color{cyan}\emph{Rifat: #1}}\xspace}
\else
  \usepackage[disable]{todonotes}
  \newcommand{\didar}[1]{}
  \newcommand{\muneera}[1]{}
  \newcommand{\fra}[1]{}
  \newcommand{\rifat}[1]{}
\fi

%% file: Sections/1_Introduction.tex
\section{Introduction}
\label{sec:intro}

The rapid proliferation of Artificial Intelligence (AI) technologies has brought both remarkable advancements and significant challenges. Among these challenges, the issue of diversity and inclusion (D\&I) has attracted considerable attention \cite{zowghi2023diversity}. Addressing D\&I in AI is crucial for several reasons: it reduces biases, increases fairness, enhances creativity, and prevents harmful societal impacts \cite{shams2023ai}. However, despite these pressing needs, D\&I considerations are often overlooked in the design, development, and deployment of AI, resulting in unintended consequences and many AI incidents.

In recent years, reported real-world AI incidents that demonstrate discrimination and bias show the necessity of integrating D\&I principles in AI applications. Instances such as Google Images misrepresenting women's job roles \cite{url1} and Google Photos mistakenly categorizing images of African Americans inappropriately \cite{url2} highlight the built-in biases of AI and the complex ethical problems that come with it. Even Tinder Plus, a popular platform, encountered issues when it implemented a biased personalized pricing algorithm that disproportionately charged users over 30 and gay and lesbian users aged 18-29 \cite{url3}. Incidents of facial recognition errors leading to wrongful arrests \cite{url5}, AI hiring tools biased against females \cite{url4}, and medical algorithms that prioritize white patients over black patients \cite{url6} clearly indicate deep-rooted biases in AI systems. Historical biases in motion capture data, which predominantly favored able-bodied male subjects \cite{url7}, further exhibit the systemic exclusion of the disabled present in AI applications. Recent incidents, such as OpenAI's ChatGPT displaying gender bias in recommendation letters \cite{url8}, only underscore the pressing need for embedding D\&I principles in AI. Beyond sex, age, or race, many AI incidents occur based on different diversity attributes such as ethnicity, language, religion, nationality, disability, culture, socio-economic status, geographic location and so on.

Analyzing AI incidents through a D\&I lens becomes critical for several compelling reasons. Firstly, it enables us to identify and understand the underlying causes of biases and discriminatory practices in AI systems. Recognizing these causes is vital for developing strategies that mitigate such biases, ensuring that future AI technologies are fairer and more equitable. Inspired by the frequent monitoring and maintenance of systems like the Black-box flight data recorder in aviation industry \cite{sear2001arl}, we posit that it is essential to adopt a similar approach in dealing with AI systems. By learning from the past mistakes, we can enhance the reliability and trustworthiness of AI systems and prevent similar incidents from occurring in the future. 

Despite the need of investigating D\&I-related AI incidents, to the best of our knowledge, no research has been conducted to identify D\&I related AI incidents, nor to propose strategies to avoid them. An important question to ask is: What is Diversity and Inclusion in AI? Zowghi et al. defined D\&I in AI with ``inclusion of humans with diverse attributes and perspectives in the data, process, system, and governance of the AI ecosystem'' \cite{zowghi2023diversity}. They proposed a 5-pillar framework in the AI ecosystems to propose a holistic and sociotechnical approach to D\&I in AI consideration. Shams et al. conducted a systematic literature review (SLR) to explore the challenges and corresponding solutions to address D\&I in AI and enhance D\&I practices by AI \cite{shams2023ai}. They utilised that 5-pillar framework of Zowghi et al. to structure and present the results of their SLR. Another research has emphasized making AI diverse and inclusive to better meet everyone's needs, respect rights, and match current societal values \cite{fosch2022diversity}. Chi et al. reported that big companies like Google, Microsoft, and Salesforce, while discussing diversity and inclusion in their AI ethics rules, are focusing more on technical aspects and that of fairness \cite{chi2021reconfiguring}. Another recent study identified that most guidelines for AI mainly focus on fairness, justice, and discrimination, while ignoring diversity, equity, and inclusion \cite{cachat2023diversity}. None of these  recent research articles has focused on D\&I issues in the reported AI incidents.

To fill this gap, we have worked on establishing a set of criteria to effectively identify D\&I-related AI incidents. For this purpose, we manually analyzed the AI incidents from two databases (AIID\footnote{ \href{https://incidentdatabase.ai/}{\url{https://incidentdatabase.ai/}}} and AIAAIC\footnote{ \href{https://www.aiaaic.org/aiaaic-repository}{\url{https://www.aiaaic.org/aiaaic-repository}}}), to classify the incidents into three categories: ``related to diversity and inclusion'', ``not related to diversity and inclusion'', and ``more information required'' to decide. We developed a decision tree to investigate the diversity and inclusion issues in AI incidents. To validate our categorization and the decision tree, we conducted card sorting exercise and focus group discussions with artificial intelligence/machine learning (AI/ML) researchers and practitioners who also have sufficient knowledge about D\&I. Finally, we have designed, populated, and made publicly available repository of D\&I-related AI incidents, intended for use for future research. The key contributions of this research are:

\begin{enumerate}
    \item \textbf{Identification of D\&I-Related AI Incidents:} We established criteria and explored two AI incident databases (AIID and AIAAIC) to categorize incidents based on their relevance with D\&I. We also provided the reason behind our categorization and the associated diversity attributes for each incident.
    \item \textbf{Development of Analytical Tools:} We created a decision tree to investigate D\&I issues in AI incidents and validated through participatory activities.
    \item \textbf{Public Repository Creation:} We developed a publicly accessible repository of D\&I-related AI incidents. This resource provides valuable insights for researchers, developers, and policymakers, guiding the responsible development of AI systems.
    \item \textbf{Assistance of Exploring Underlying Causes of D\&I-Related AI Incidents:} Our research sheds light on the causes of AI incidents related to D\&I, that could assist in proposing potential strategies to avoid such incidents in future.
    \item \textbf{Promotion of Responsible AI Practices:} The overarching goal of this research is to enhance awareness among researchers and practitioners about the critical need to embed D\&I principles comprehensively in AI systems, aiming for more inclusive and equitable technology development and application. Our repository serves as a crucial resource for researchers, developers, and policymakers, offering valuable lessons and guidelines for the responsible development and deployment of AI systems.
\end{enumerate}

\textbf{Paper Organization.} Section \ref{sec:background} discusses the background of this research and the related work. In Section \ref{sec:methodology}, we explain our research methodology. Section \ref{sec:results} reports the results of this study which we discuss in Section \ref{sec:discussion}. Section \ref{sec:ttv} discusses the possible threats to validity of this research. Finally, we conclude our research with possible future research directions in Section \ref{sec:conclusions}.

%% file: Sections/2_Background.tex
\section{Background and Related Work}
\label{sec:background}

The evolution of artificial intelligence (AI) \cite{singari2022contemporary} has permeated many domains, including health \cite{rajpurkar2022ai}, education \cite{zhai2021review}, transportation \cite{abduljabbar2019applications}, and law \cite{atkinson2020explanation}, necessitating the development of ethical and responsible systems. Discrimination can be embedded into AI systems through various avenues such as data, design, implementation, and the absence of adequate legal frameworks \cite{tao2022research}. Data used to train AI models may contain biases, leading to algorithmic discrimination \cite{von2021discriminatory}. The design of AI algorithms can inadvertently perpetuate discriminatory outcomes, even when human prejudices are intended to be eliminated. Furthermore, the implementation of AI systems without proper checks and balances can result in discriminatory decisions \cite{ferrer2021bias}. The lack of comprehensive legal frameworks to regulate AI and prevent discriminatory practices poses a significant challenge \cite{wachter2021fairness}. To address these issues, a multidisciplinary approach integrating legal and technological perspectives is crucial to develop fair and unbiased AI systems that comply with existing antidiscrimination laws \cite{arnanz2023creating}. According to Zhou et al., the widespread application of AI in various domains makes it imperative to align its operational principles with ethical standards \cite{zhou2020survey}. This necessitates the establishment of guidelines and principles to ensure such systems are unbiased, trustworthy, and fair to all. 

Principles of Diversity and Inclusion aim to tackle the challenges of bias and discrimination in society. Embedding D\&I principles into the processes of designing, developing, and deploying AI systems is crucial to achieving equity and fairness \cite{zowghi2023diversity}. An important aspect of integrating D\&I into AI involves the identification and analysis of D\&I-related AI incidents. These incidents expose the underlying biases and discrimination embedded within AI systems. Identifying these incidents enables us to understand their causes and develop strategies to mitigate them, thereby enhancing the fairness and inclusivity for future AI technologies.

\subsection{Diversity and Inclusion in AI}
D\&I in AI is gaining increasing attention in research and practice. To achieve trustworthy AI, the importance of embedding diversity and Inclusion throughout the AI system development life cycle has been emphasized.  \cite{zowghi2023diversity}. While D\&I and fairness are distinct concepts, fostering diversity can lead to fair outcomes, particularly in information access systems like recommendation systems and search engines \cite{porcaro2023fairness}. Scholars highlight the risks of AI systems perpetuating existing inequalities, underscoring the need for responsible AI development that incorporates diversity, equity, and inclusion (DEI) principles and practices \cite{cachat2023diversity}. Guidelines for AI increasingly advocate for DEI principles, emphasizing the importance of addressing DEI risks through actions that influence AI actors' behaviors and awareness \cite{cachat2023diversity}.

In order to have a comprehensive understanding of diversity in AI, it is vital to acquire an understanding of different diversity attributes (e.g., gender, age, ethnicity, race, socio-economic status, nationality, religion etc.) that necessitate careful consideration within AI systems. Zowghi et al. defined diversity attributes as ``known facets of diversity, including (but not limited to) the protected attributes in Article 26 of the International Covenant on Civil and Political Rights (ICCPR) \cite{australian2014quick}, as well as race, colour, sex, language, religion, national or social origin, property, birth or other status, and inter-sections of these attributes'' \cite{zowghi2024ai}. AI's impact on various diversity attributes looks into how AI systems can either include or exclude diverse groups such as women, LGBTQI+ individuals, different races, age groups, and people with different abilities, depending on how the data is selected, how the AI is trained, and how it is ultimately used \cite{soraa2023ai}. While several studies have explored gender diversity in AI, numerous other dimensions of diversity attributes have been largely overlooked in AI research \cite{shams2023ai}. Similarly, in practice, AI systems like facial recognition, voice recognition, and prediction systems have a high probability of impacting diversity attributes such as race, gender, dialects etc. For example, studies have shown that automated face recognition systems are significantly impaired by demographic attributes, that leads to a significant decrease in face recognition performance \cite{mah2023assessing}.

\subsection{AI Incidents Related to Diversity and Inclusion}
Since D\&I in AI is a new and growing field of study, there is a paucity of research on this topic. This gap has led to a lack of awareness in implementing D\&I in AI systems, consequently resulting in various AI incidents that could be attributed to the violation of D\&I principles. There are publicly available AI incident databases such as AI Incident Database (AIID), AI, Algorithmic, and Automation Incidents and Controversies (AIAAIC), and OECD AI Incident Database\footnote{ \href{https://oecd.ai/en/catalogue/tools/ai-incident-database}{\url{https://oecd.ai/en/catalogue/tools/ai-incident-database}}}.

A number of recent studies have focused on AI incident databases for various purposes. For example, a recent research emphasized the importance of an AI incident database for recording and examining real-world AI failures, thus preventing recurring mistakes and ensuring AI's societal benefits \cite{mcgregor2021preventing}. Wei et al. presented a detailed analysis of real-world AI ethical issues, drawn from the AI Incident Database, identifying 13 prevalent application areas and 8 forms of ethical issues with the aim to provide AI practitioners with a practice-oriented guideline for ethical AI deployment \cite{wei2022ai}. Similarly, another study addressed challenges to engineering trustworthy AI by analyzing 30 real-world incidents of trust loss from the AI incident database, offering practical recommendations to be incorporated into the development cycle in AI systems \cite{stanley2023exploring}. Feffer et al. suggested the AI incident database as an educational tool, highlighting its role in a study that enhanced students' understanding of AI harms and the requirement for safe and responsible AI \cite{feffer2023ai}. A recent research utilized public AI incident databases to assess reporting techniques, aiming to enhance incident documentation, thus contributing to safer, fairer AI development \cite{turri2023we}.

While several recent studies have used AI incident databases for a variety of research objectives, there appears to be a noticeable gap in current state of the art on investigating D\&I issues in AI incidents. To the best of our knowledge, no research has been undertaken so far that specifically targets D\&I issues within AI incidents. This deficit extends to inquiries into the causes of D\&I-related AI incidents, as well as the formulation of strategies that could potentially prevent such incidents. To mitigate this gap, we undertook a study utilizing two AI incident databases (AIID and AIAAIC) with the aim of identifying AI incidents that are related to diversity and inclusion (D\&I). Part of our initiative was to also propose methods for such identification. Further, we have developed a publicly accessible repository of D\&I-related AI incidents. This under explored area of AI research, which combines D\&I considerations with incident analysis, hence presents a significant opportunity for future study, and could contribute greatly to our understanding and management of D\&I issues in AI.

%% file: Sections/3_Methodology.tex
\section{Methodology}
\label{sec:methodology}

With the aim to identify AI incidents related to diversity and inclusion, we formulated the following two research questions.

    \indent \textit{\textbf{RQ1.} How can we identify if an AI incident is related to diversity and inclusion issues?}\\ 
    \indent \textit{\textbf{RQ2.} To what extent do the existing AI incidents related to diversity and inclusion issues?}

\begin{figure*}[!htbp]
            \centering
            \includegraphics[width=0.9\textwidth]{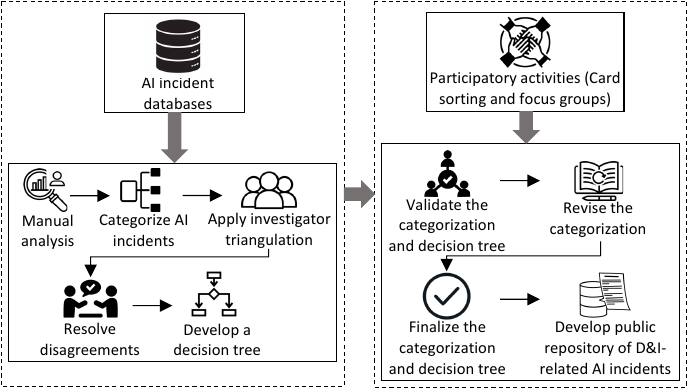}
            \caption{An overview of the research method}
            \label{fig:method}
    \end{figure*}

We conducted a mixed-methods empirical study to answer the research questions. We collected data from two public online AI incident databases, and through a card sorting exercise and two focus group discussions with artificial intelligence/machine learning (AI/ML) researchers and practitioners, who are enthusiast of D\&I. Figure \ref{fig:method} shows an overview of our research method. As this study worked with humans, ethics approval was acquired from our organization's Human Research Ethics Committee on 19/03/2024.

\subsection{Data Collection}
We collected data from AI incident databases, card sorting exercise and focus groups.

\subsubsection{AI Incident Databases}
\leavevmode
\newline
We collected AI incidents from two publicly available databases, AI Incident Database (AIID) and AI, algorithmic, and automation incidents and controversies (AIAAIC). These databases are dedicated archives, indexing the collective history of potential harms that have transpired in real-world scenarios as a result of deploying AI systems. 
We collected the data from AIID in July 2023 and from AIAAIC in March 2023. There were 551 AI incidents in AIID database and 575 incidents in AIAAIC. The AIID database includes the titles and summaries of the incidents including some news links, dates of the incidents, and information about the alleged/harmed/nearly harmed parties. On the other hand, AIAAIC includes the titles and summaries of the incidents along with the incident year, country, sector (e.g., media, sports, health), operator (e.g., Google, Microsoft), developer, system, technology (e.g., facial recognition, natural language processing) and corresponding news links. We manually analyzed the incidents from D\&I lens and categorized them under three groups: Related to D\&I (R), Not related to D\&I (NR) and More information required (MIR). We also mentioned the reason supporting our determination for each incident and identified the diversity attributes for the D\&I-related AI incidents.


\subsubsection{Participants of Card Sorting and Focus Groups}
\leavevmode
\newline
We conducted card sorting to validate our categorization process of AI incidents with their relevance with D\&I issues, as card sorting is a reliable, fast, and inexpensive approach that provides comprehensive understanding on the subject matter \cite{spencer2004card}. Additionally, as focus groups are a valuable validation technique used in various research fields \cite{lichtenberg2000prospective}, \cite{fisk2024multi}, \cite{subaeki2022assessing}, we also conducted focus groups to validate our categorization process and the decision tree.

Initially, we formulated a specific set of recruitment criteria, facilitating the identification of suitable participants for the card sorting and focus groups. The main criterion to select the participants was finding practitioners who are AI researchers or AI practitioners, or information system and computer science researchers who have knowledge about diversity and inclusion. We shared our project objectives and participant recruitment criteria through social media (LinkedIn, Twitter (now X), Facebook) to invite participants in our study. We also invited potential participants through emails. Seven participants agreed to join the activities. All of them joined the card sorting, however, one participant did not attend the focus group. Therefore, we conducted two focus groups with three participants in each group.

At first, they were given participant information sheet to know the details of the project and their roles. It also included the risks and benefits of the project, withdrawal of participation from the project, confidentiality of their identities, data management, and contact details of the research team and human ethics research approval committee. Later, the participants were provided with a consent form and asked to give their consents by signing and returning the consent form.


\subsubsection{Protocol of Card Sorting}
\leavevmode
\newline
We conducted card sorting with seven participants. We prepared 10 cards for each participant, where each card contained one AI incident. We selected 5 incidents from AIID database and another 5 incidents from AIAAIC database that were particularly thought-provoking and represented different categories (R, NR, MIR) according to our manual analysis. As manual analysis takes time, analyzing more than 10 incidents in a participatory activity was difficult. This activity was conducted in our workplace that took approximately 60 minutes.

This method consisted of three parts. The first part was an icebreaker; the principal investigator of this study met the participants before data collection starts and spent some time with them chatting to help put them at their ease. The second part covered explaining our research, its objectives, expected outcomes, possible benefits and risks of this research, and the strategies adopted to ensure the confidentiality of the collected data. For ethical compliance, we provided the participants with a participant information sheet and sought their signed consent. The last part was the closed card sorting \cite{kerr2010information}, where the participants were given the cards containing AI incidents. They were asked to group the incidents under three categories based on the relevance of the incidents with D\&I issues. 
Each card was numbered so that they can be provided in the same order to everyone. The participants were asked to analyze each card and categorize it based on the given categories. At this stage, they were asked to make their own decision without any discussion with other participants. They were also asked to write the reasons behind their labelling decision for each incident.


\subsubsection{Protocol of Focus Groups}
\leavevmode
\newline
As focus groups are a valuable validation technique used in various research fields \cite{lichtenberg2000prospective}, \cite{fisk2024multi}, \cite{subaeki2022assessing}, we conducted two focus groups with three participants in each focus group to validate our categorization process and the decision tree. We arranged the focus groups at our workplace. The participants were provided a comfortable environment with round seating arrangements. 

Each focus group was divided into two parts. In the first part, the participants were encouraged to discuss about the cards with AI incidents and the way they categorized the cards. In the second part, we provided the decision tree to the participants. We described how we developed the decision tree and how the decision tree works to identify the diversity and inclusion issues in AI incidents. Afterwards, the participants were encouraged to criticize the decision tree and provide recommendations on how to improve it based on their knowledge on the card sorting and the first part of the focus group discussion. We Asked the following questions to facilitate the focus groups.

\begin{itemize}
    \item \textbf{Questions on Card Sorting Exercise:} How did you categorize the given AI incidents?, Have you faced any difficulties in categorizing them?, When you categorized an incident under ``More information required'', what information do you think should be provided to categorize them?
    \item \textbf{Questions on the Decision Tree:} Do you think the decision tree is aligned with your categorization? Do you think any step of the decision tree is not clear enough to understand? Do you have any comments/concerns/recommendations on the decision tree?
\end{itemize}

Given the nature of focus groups to encourage participants to have an in-depth discussion, these sessions provided an invaluable opportunity to extract the required data and validate our proposed decision tree. We arranged the two focus groups on two different days. The first focus group took 43 minutes and the second one took 47 minutes. We recorded the focus group discussions after obtaining participants' consent.


\subsection{Data Analysis}
Figure \ref{fig:method} shows the overview of the data analysis. The analysis was divided into the following steps to ensure maximum clarity and precision.

\textbf{Collection and Initial Analysis of AI Incident Data.} We selected 551 incidents from the AI Incident Database (AIID), timestamped on 24/07/2023. The first author of this paper manually analyzed the incidents to explore their correlation to D\&I. The incidents were categorized into three distinct groups based on their relation to D\&I. 

\textbf{Investigator Triangulation.} To promote confidence in the initial analysis, investigator triangulation was applied to the first 100 AI incidents. Another researcher from our team was asked to perform an independent analysis to the incidents. We experienced 26 disagreements in categorization. However, they were resolved after having a discussion between the two investigators, allowing for a deeper understanding of the complex dynamics inherent within each case. By investigating these incidents from multiple perspectives, we increased confidence in the validity of our findings.

\textbf{Re-Analyzing Incident Data and Decision Tree Development.} The first author manually analyzed all the 551 incidents once again taking into account the insights obtained from the discussion. As a result the decision for 45 incidents were changed. The first author also analyzed the first 50 incidents from AIAAIC database with the knowledge developed from the analysis of AIID and the discussion with the second investigator. This analysis and discussion helped in the development of the initial decision tree (V1) to identify D\&I-related in AI incidents (see Figure \ref{fig:decision_tree_v1} in Appendix \ref{appendix_A}). Based on the conditions on how we categorized an incident to check the relevance of an incident with D\&I issues, we developed this decision tree. Similar to our manual categorization, our proposed decision tree also provides three categories of AI incidents based on their relevance with D\&I: related to D\&I, not related to D\&I and more information required.

\textbf{Updating Decision Tree.} The preliminary analysis and the decision tree were shared with the co-authors of this paper who are also experts in D\&I in AI. After several rounds of discussions and iteration, we updated the initial decision tree (V2) accordingly guided by the improved understanding of the incidents (see Figure \ref{fig:decision_tree_v2} in Appendix \ref{appendix_A}).

\textbf{Organizing Participatory Activities.} We organized card-sorting and focus group discussions on April 2024 with 7 and 6 participants respectively to validate our categorization and the decision tree. In the card sorting exercise, all of the participants' categorizations were exactly similar to our categorization for three incidents. For five incidents, their categorizations closely aligned with our own. 
Different opinions came from different participants for the remaining two incidents. Despite divergent perspectives, our focus group discussions were insightful and productive, leading to enhanced clarity in our understanding of the varied perceptions. Both focus group discussions were recorded and transcribed for facilitating detailed data analysis.

\textbf{Finalizing Categorization and Decision Tree.} After conducting the card-sorting and focus groups, and subsequent analysis of qualitative data from participatory activities, the first author conducted a manual analysis of all 551 incidents for the third time and finalized the incident categorization. The decision for 17 incidents were updated again. Following the first focus group discussion, we updated the decision tree (V3) (see Figure \ref{fig:decision_tree_v3} in Appendix \ref{appendix_A}). After the second focus group discussion and the revision of the incident categorization, slight modifications were done to the decision tree once again (V4) (see Figure \ref{fig:decision_tree_v4} in Appendix \ref{appendix_A}). 
Our process stopped when everyone was satisfied with the final decision tree.

\textbf{Applying the Decision Tree on Another Database.} The first author applied the decision tree manually for the first 310 incidents from AIAAIC database and identify the D\&I-related AI incidents. However, we will analyze all the incidents from AIAAIC and populate our repository with new incidents in future.

\textbf{Public Repository Creation.} Finally, we created a public repository of D\&I-related AI incidents for both AIID and AIAAIC. This open-source repository will serve as a valuable resource for other researchers in this field. However, the repository requires continuous updates as new incidents are constantly being added to the incident databases. 

%% file: Sections/4_Results.tex
\section{Results}
\label{sec:results}

This section presents the results derived from the analysis of the AI incident database. The results also reflects the analytical overview of the card sorting exercise and focus group discussions carried out with AI/ML researchers and practitioners.

\subsection{RQ1: Identifying D\&I Issues in AI Incidents}
\begin{figure*}[!htbp]
            \centering
            \includegraphics[width=0.9\textwidth]{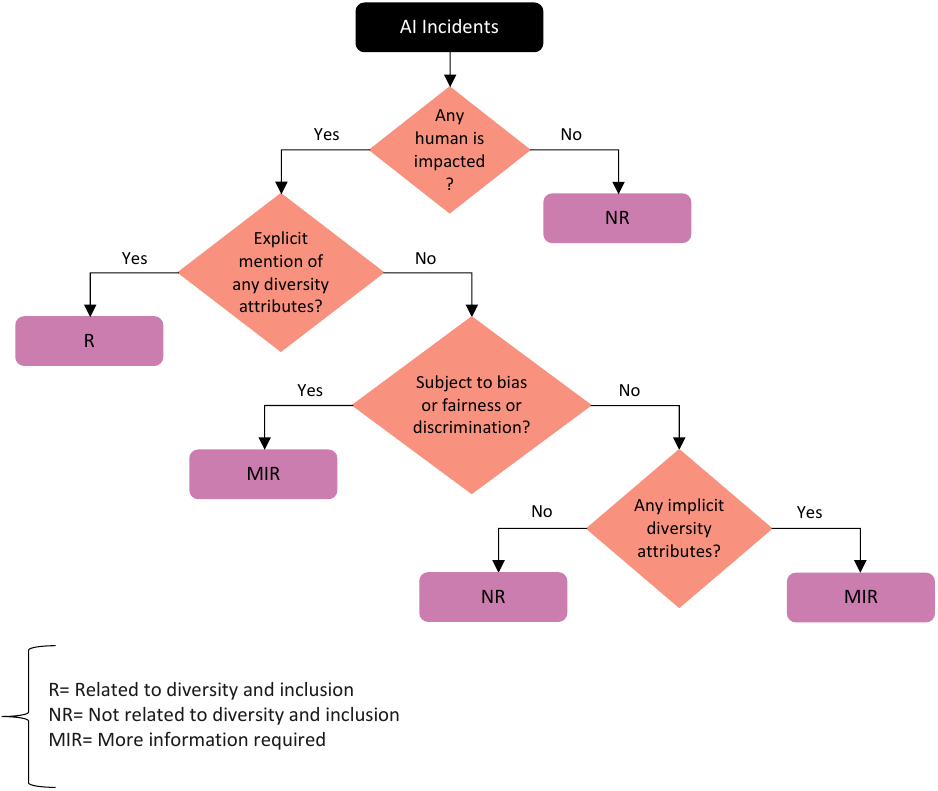}
            \caption{Decision tree to detect D\&I-related AI incidents}
            \label{fig:decision_tree}
    \end{figure*}

Figure \ref{fig:decision_tree} shows the results of RQ1. To identify Diversity and Inclusion (D\&I) related AI incidents, we developed a decision tree after a rigorous analysis. 

Our proposed decision tree has four conditions. The first condition checks if any human is directly or indirectly impacted by the AI incident. The impact can be physical, psychological, financial and so on. If the response is negative, the incident is not related to D\&I. On the other hand, a positive response directs the process to the second condition, which examines whether the AI incident explicitly mentions about any diversity attributes, including but not limited to 
gender, sex, sexual orientation, age, skin tone, race, ethnicity, religion, language, literacy, disability, neurodiversity, facial features, physical features, nationality, accent, culture, geographic location, socio-economic status, and political ideology. If the answer is a `yes', the incident is clearly related to D\&I issues. However, for the answer `no', the incident is scrutinized under the third condition, whether this AI incident is subject to bias or violates fairness or results in discrimination. As inferred from the preceding condition, the incident does not explicitly link to any diversity attributes. Still, if the elements of bias, fairness, or discrimination are present, it necessitates additional information before drawing a definitive conclusion about its relevance or irrelevance to D\&I. A negative determination on this condition points towards a lower likelihood of direct D\&I applicability. Nonetheless, a final check must be conducted on whether the incident posses any implicit diversity attributes. For example, if the impacted parties 
represent any diverse communities such as a specific gender or a particular age group 
or if the name(s) of the impacted parties highlight any diversity attributes, such as association with a particular religious faith, then we need more information to determine if the incident occurred due to the violation of these diversity attributes. Therefore, a positive outcome needs a more investigation to ascertain any potential correlation with D\&I issues, whereas a negative implies an absence of a D\&I implication.

    \begin{figure*}[!htbp]
            \centering
            \includegraphics[width=1.03\textwidth]{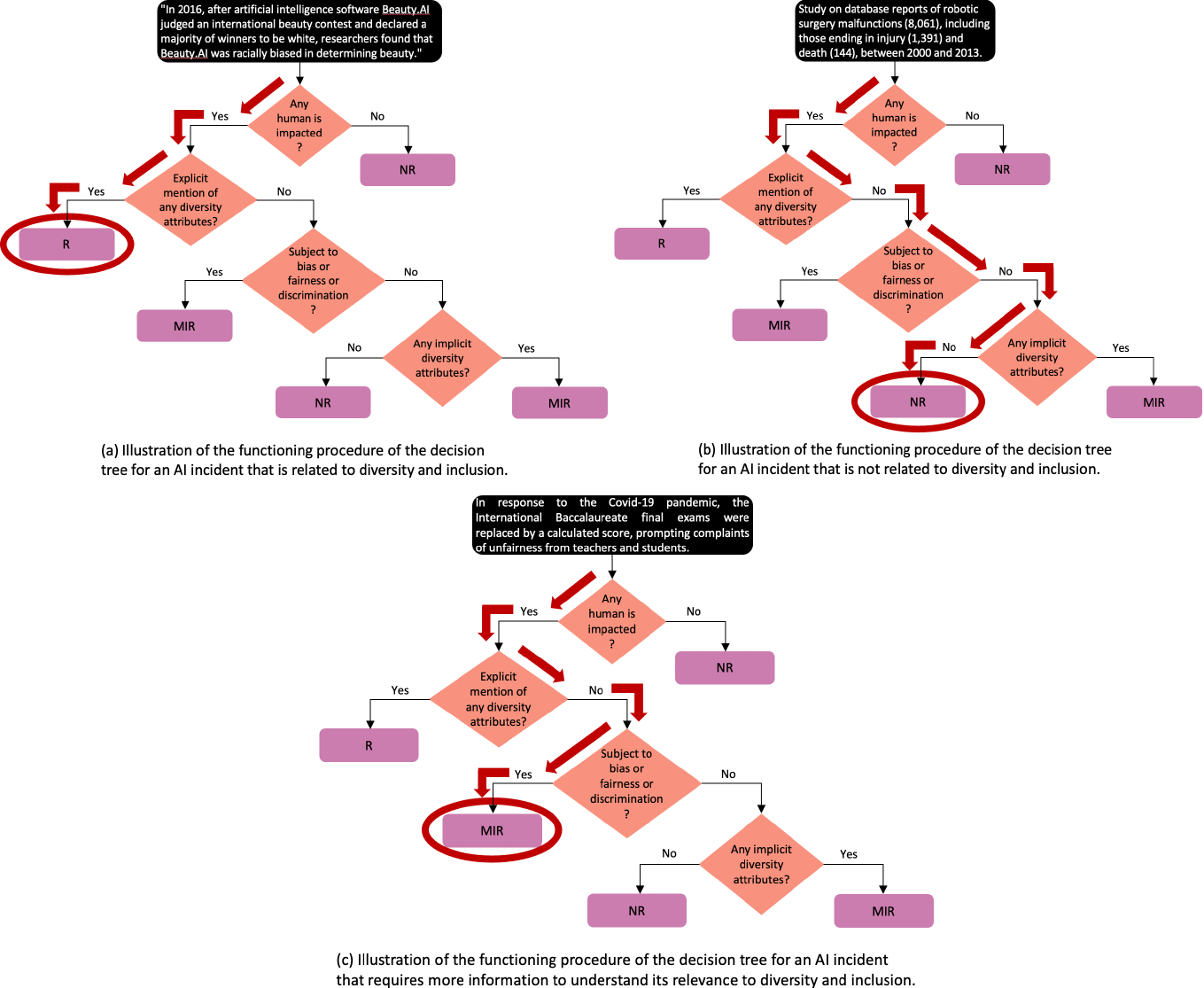}
            \caption{Illustration of the functioning procedure of the decision tree to understand its relevance to diversity and inclusion with some examples.}
            \label{fig:example}
    \end{figure*}

Figure \ref{fig:example} presents several illustrative examples of AI incidents to elucidate the procedure of identifying their association with D\&I issues. Figure \ref{fig:example} (a) demonstrates the trajectory of an AI incident within the decision tree that ultimately leads to the conclusive result of ``related to D\&I''. The AI incident states:
    \begin{itemize}
        \item[\faCommentsO] \textit{``In 2016, after artificial intelligence software Beauty.AI judged an international beauty contest and declared a majority of winners to be white, researchers found that Beauty.AI was racially biased in determining beauty.''}
    \end{itemize}
    
As humans are impacted through this AI incident, the process moves forward to the second condition and examines whether this incident is explicitly related to any diversity attributes. Since this incident clearly has racial undertones, the decision pathway moves forward to the left, concluding that the incident has relevance to D\&I issues. Similarly, Figure \ref{fig:example} (b) shows an example of an AI incident that is not related to D\&I issues.
    \begin{itemize}
        \item[\faCommentsO] \textit{``Study on database reports of robotic surgery malfunctions (8,061), including those ending in injury (1,391) and death (144), between 2000 and 2013.''}
    \end{itemize}

The above-mentioned AI incident evidently caused harm on humans, therefore, the procedural control moves towards the left side of the decision tree. Here, the incident is examined for any direct alignments with diversity attributes. As, the incident does not explicitly mention any diversity attributes, the process shifts towards the right to examine if it represents any bias, breaches fairness or carries discriminatory implications. In absence of clear indications, the control moves towards the next condition and tests whether the incident has any implicit diversity attributes. Since the outcome remains negative, the final determination categorizes the incident as ``not related to D\&I''. Similarly, Figure \ref{fig:example} (c) demonstrates the navigation of another AI incident through the decision tree, leading to an outcome of ``more information required''.
    \begin{itemize}
        \item[\faCommentsO] \textit{``In response to the Covid-19 pandemic, the International Baccalaureate final exams were replaced by a calculated score, prompting complaints of unfairness from teachers and students.''}
    \end{itemize}

The indicated incident certainly impacted humans, prompting the control to proceed towards the left of the decision tree for assessing any direct relevance with diversity attributes. When no explicit association is uncovered concerning diversity attributes, the control advances to the next condition and examines whether the incident exhibits bias, violates fairness, or results in discrimination. As this incident clearly breaches fairness, the process moves to the left, indicating that further information is required to determine the incident's relevance to D\&I.


\subsection{RQ2: Extent to Which the Existing AI Incidents are Related to D\&I issues}
Of the 551 AI incidents extracted from the AIID database, our analysis identified 189 AI incidents related to D\&I issues (see Table \ref{tab:rq2_extent_AIID}). 80 incidents require further information to establish their correlation with D\&I issues. The rest of the 282 incidents are not related to D\&I. 
Similarly, from AIAAIC database, we identified almost half of the incidents are related to D\&I issues, 144 D\&I-related AI incidents among the 310 incidents we analyzed (see Table \ref{tab:rq2_extent_AIAAIC}). 50 incidents require more information to make determination. The rest of the 116 incidents are not related to D\&I issues.

\begin{table}
    \centering
\begin{tabular}{|p{3cm}|c|c|}
    \hline 
         \textbf{Status}&  \textbf{No. of AI incidents}& \textbf{Percentage}\\ \hline 
         Related to D\&I&  189& 34.3\%\\ \hline 
         Not related to D\&I&  282& 51.18\%\\ \hline 
         More information required&  80& 14.52\%\\ \hline
    \end{tabular}
    \caption{The extent to which the existing AI incidents are related to D\&I issues from AIID Database}
    \label{tab:rq2_extent_AIID}
\end{table}

\begin{table}
    \centering
\begin{tabular}{|p{3cm}|c|c|}
    \hline 
         \textbf{Status}&  \textbf{No. of AI incidents}& \textbf{Percentage}\\ \hline 
         Related to D\&I&  144& 46.45\%\\ \hline 
         Not related to D\&I&  116& 37.42\%\\ \hline 
         More information required&  50& 16.13\%\\ \hline
    \end{tabular}
    \caption{The extent to which the existing AI incidents are related to D\&I issues from AIAAIC Database}
    \label{tab:rq2_extent_AIAAIC}
\end{table}

We developed the first version of D\&I-related AI incidents repository based on the AIID and AIAAIC databases \cite{shams_rifat_ara_11639709}. Each record in this repository encompasses details such as the incident ID, title, description, date, alleged deployer of AI system, alleged developer of AI system, and alleged harmed or nearly harmed parties and so on. Each incident ID is linked to a corresponding article for a more detailed account of the incident. Further, in the repository, we added the status in relation to D\&I, diversity attributes such as age, gender, ethnicity, race and so on, and the reason supporting our determination. We have outlined two categories under the ``status'' column to categorize the incidents: related to D\&I (R) and more information required (MIR). Let us consider an example.
    \begin{itemize}
        \item[\faCommentsO] \textit{``Google's Perspective API, which assigns a toxicity score to online text, seems to award higher toxicity scores to content involving non-white, male, Christian, heterosexual phrases.''}
    \end{itemize}

We labeled the above-mentioned AI incident as being ``related to D\&I''. The reason for our decision was ``Google's API provided higher toxicity scores to non-white, male, Christian, heterosexual phrases. This is a clear breach of D\&I''. The diversity attributes involved in this incident encompassed race, gender, sexual orientation, and religion. Similarly, we have another instance of an AI incident stating:
    \begin{itemize}
        \item[\faCommentsO] \textit{``An algorithm used to rate the effectiveness of school teachers in New York has resulted in thousands of disputes of its results.''}
    \end{itemize}

The aforementioned AI incident is classified as ``more information required''. Our reasoning for this decision was articulated as such: ``The situation might involve bias or fairness problems if the algorithm was trained with skewed data or overlooked factors like class size or student backgrounds, possibly leading to unfair teacher ratings. However, disputes might stem from disagreement with outcomes, not algorithm unfairness or discrimination, so more details are needed to confirm if the algorithm is biased''.

    \begin{figure*}[!htbp]
            \centering
            \includegraphics[width=0.8\textwidth]{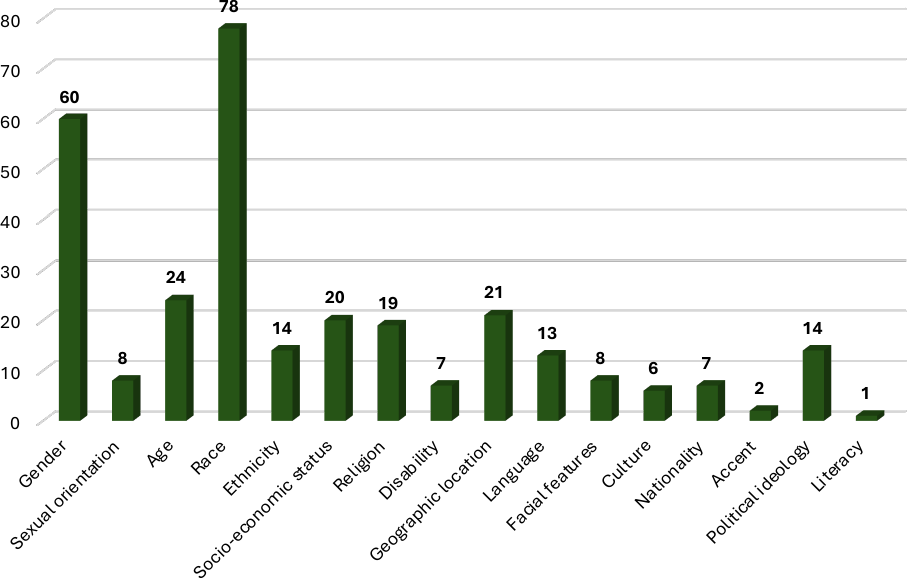}
            \caption{The ratio of diversity attributes in AI incidents from AIID database}
            \label{fig:attributes_AIID}
    \end{figure*}

Figure \ref{fig:attributes_AIID} shows the proportion of 16 different diversity attributes implicated in AI incidents correlated with D\&I from the database, AIID. The majority of these AI incidents associated with D\&I arose from discrimination based on `race'. In total, `race' is directly linked to 78 distinct incidents. The second highest diversity attributes is `gender', which is indicated in 60 AI incidents. Substantial proportions of AI incidents are also related to attributes such as `age', `geographic location', `socio-economic status', and `religion', associated with 24, 21, 20, and 19 incidents, respectively.

        \begin{figure*}[!htbp]
            \centering
            \includegraphics[width=0.8\textwidth]{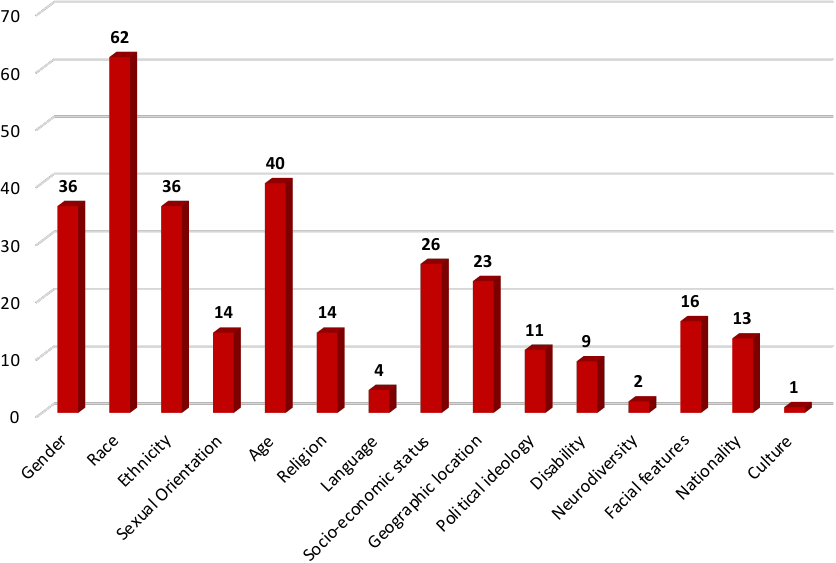}
            \caption{The ratio of diversity attributes in AI incidents from AIAAIC database}
            \label{fig:attributes_AIAAIC}
    \end{figure*}

Similarly, the ratio of 15 different diversity attributes for the D\&I-related AI incidents from AIAAIC database are shown in Figure \ref{fig:attributes_AIAAIC}. Similar to AIID, this database also has the maximum incidents based on `race'; 62 incidents occurred due to racial bias. The second highest diversity attribute for AIAAIC is `age', associated with 40 incidents. Bias based on `gender' and `ethnicity' linked to same number of incidents (36). Additionally, significant number of incidents occurred due to bias on `socio-economic status' and `geographic location', 26 and 23 incidents respectively.

%% file: Sections/5_Discussion.tex
\section{Discussion and Implications}
\label{sec:discussion}

This section discusses the findings of this study as well as the implications for research and practice.

\textbf{First Step Towards Identifying Cause of D\&I-related AI Incidents.} Identifying D\&I-related AI incidents is the first step towards further investigations on these incidents to explore the underlying causes of them and identify potential strategies to avoid them. To accurately identify D\&I-related AI incidents, we iteratively constructed a decision tree to systematically assess the D\&I issues of an AI incident. This will guide further research trying to better understand the causes of these incidents. However, the decision tree can only provide preliminary insights unless it is complemented by a thorough causal analysis as well as strategies to avoid future AI incidents based on D\&I issues.

\textbf{The Role of AI in D\&I Issues.} Our research indicates a significant number of AI incidents linked to D\&I challenges. With 189 out of 551 incidents (34.3\%) from AIID and 144 out of 310 (46.45\%) from AIAAIC found as related to D\&I issues. This finding underscores the role AI plays in amplifying or generating D\&I concerns. Careful attention must be paid to D\&I during AI system design, development and deployment. Otherwise, these systems may have the potential to cause further harm due to biases inherent in their design and deployment.

\textbf{The Uncertainties Around AI and D\&I.} Despite our decision tree, considerable uncertainty remains, as 80 out of 551 incidents from AIID and 50 out of 310 incidents from AIAAIC still require additional information to classify. This shows the complexity involved in dealing with D\&I in AI, affirming the need for deeper analysis and further research.

\textbf{Necessary Information for Categorizing AI Incidents.} Due to the lack of adequate information, significant number of incidents were categorized under ``more information required''. Thus, the summaries of the AI incidents need to be more informative and comprehensive to facilitate the determinations regarding any potential connections to D\&I issues. If data privacy rules allow it, the summaries should incorporate the names of the accused or harmed parties that might signal elements of diversity, such as religion, ethnicity, race, gender, or age. The objective is to determine whether the incident was unintentional or incited by biases. Equally, if an AI system wrongly impacted an individual, it is crucial to establish whether any biases played a role in this incident.

Bias in AI systems can manifest through various pathways, significantly impacting the fairness and reliability of the outcomes. One primary source of bias is the training data \cite{kavitha2022prediction}. Another critical aspect is the potential bias within the models or algorithms \cite{shin2023data}. Design choices, such as the selection of features or the weighting mechanisms in algorithms, can introduce or exacerbate biases \cite{pyatkin2023design}. Furthermore, there are instances where issues within AI systems arise from genuine bugs independent of any biases. These bugs could result from coding errors, inadequate system testing, or unforeseen interactions within the AI components, impacting the model's functionality without being inherently related to diversity and inclusion concerns \cite{aggarwal2023fairness}. Therefore, while biases are often linked to D\&I, it is important to recognize that not all problems in AI systems come from these biases. Some could be purely technical defects, necessitating comprehensive debugging and optimization processes. However, if there is evidence of bias, discrimination, or unfairness, the next step is to pinpoint if these are grounded in D\&I concerns. If the incident is linked to D\&I issues, it is necessary to determine whether it was the result of AI error or human interference operating via the AI platform.

\textbf{Regular Updates of the Repository.} Developing a public repository of the categorized D\&I-related AI incidents is important to provide insights into the implications and assist in raising awareness of D\&I issues among AI researchers and developers. As AI continues to evolve and be utilized across various domains, the frequency of incidents involving AI is also rising. Moreover, AI incident databases are regularly updated with new occurrences. Therefore, this repository should be continuously updated to remain relevant and beneficial.

\textbf{Significant Proportion of Racial, Gender, and Age Discrimination.} The attribute most frequently linked to discrimination is `race', with it being implicated in a total of 78 distinct incidents from AIID and 62 incidents from AIAAIC. The prominence of race-related issues in AI applications raises serious concerns about discriminatory practices and cultural biases taking root in the development phase. Following race, `gender' is the second most significant attribute from AIID with respect to discrimination, with a total of 60 documented cases. It demonstrates that AI systems might not be treating all genders equally, which can lead to serious repercussions in systems where accuracy or fairness is imperative. `Age' is also another diversity attribute which is mentioned 40 times in AIAAIC. While it is important to consider all diversity attributes we identified in this study to prevent future occurrences of D\&I-related AI incidents, particular emphasis should be placed on addressing `race', `gender', and `age' diversity within AI systems.

\textbf{Roadmap for Future Research.} Mature industrial sectors (e.g., aviation) capture their real-world failures in incident databases to inform safety improvements. AI systems currently cause real-world harm without a collective and systematic analysis of the causes of their failures. Many of these incidents are attributed to bias, unfairness, and violations of diversity and inclusion principles. As a result, companies repeatedly make the same mistakes in the design, development, and deployment of AI systems. What is lacking is a systematic approach for the thorough analysis of AI incidents to interrogate the causes, learn from mistakes, and provide actionable recommendations for all the relevant stakeholders in the AI ecosystem. 

We propose to address this challenge by developing an evidence-based framework for operationalizing diversity and inclusion in AI. This framework consists of three stages: monitoring, analysis, and investigation. The main aim is to develop methodologies to analyze and report the causes of AI incidents stemming from D\&I violations. The monitoring stage refers to the development of a portal that allows (a) others to report D\&I-related AI incidents, (b) monitors and collects D\&I-related incidents from news media and other social and public discourse, and (c) monitors the online AI incident databases for such entries. The analysis stage involves the development of sophisticated methods for deeper analysis of the D\&I-related AI incidents and preparing a structured report that contains all the pertinent information for future developers. The last stage is about development of specific criteria for investigating the incident and developing a repository of actionable recommendations.  This framework promotes transparency and trust in AI systems and supports an evidence-based AI impact assessment for the risks associated with D\&I violations in AI lifecycle. Inspired by the Black Box flight recorder concept in the aviation industry, we aim to create a similar approach in the AI ecosystem with our proposed framework to monitor, analyze, and investigate AI incidents to offer clear recommendations for improving AI safety and to continuously contribute to avoiding similar AI incidents. 

In this paper, we have presented the first step in developing our framework. Our work establishes a robust and repeatable technique for identifying D\&I-related AI incidents and creating and making an online repository of the already-reported AI incidents available for researchers and practitioners. By making the repository accessible, it is hoped that the research community can extend the analysis beyond this initial examination. It further opens the door for community contributions and validations, facilitating a more robust understanding of D\&I in AI. We plan to use the power of generative AI to automate activities in various stages of our framework for reporting and recommendations.

%% file: Sections/6_Threats_to_Validity.tex
\section{Threats to Validity}
\label{sec:ttv}

This section discusses the possible threats arising from this research based on the four validation criteria: credibility, confirmability, dependability, and transferability \cite{cruzes2011recommended}.

\textbf{Credibility.} One potential threat to credibility could stem from the fact that the AI Incident Databases (AIID and AIAAIC), which were our primary data source, may not be an exhaustive list of AI incidents related to D\&I issues. Unreported incidents, and incidents that were incorrectly reported or categorized in the database could be overlooked in the study. Furthermore, incidents reported in the database may carry a certain level of bias as they may disproportionately represent incidents from specific sectors, regions, or communities. We could overcome this threat by including other databases such as OECD AI incident database in the second version of our repository, which is our future work.

\textbf{Confirmability.} Confirmability refers to the degree of neutrality in the research findings. A significant threat to confirmability could arise from the manual analysis of incidents. Personal bias, misunderstanding or misinterpretation of information could influence our analysis and the development of the decision tree. Personal interests and preconceptions may have influenced the assessments and categorizations of incidents. Similarly, the decision-tree may have been influenced by the investigators' understanding and interpretation of D\&I issues. However, investigator triangulation and focus groups have been conducted to minimize potential bias. Furthermore, the investigators come from diverse ethnic, cultural, racial, and age groups, which enhances their perspective on diversity and inclusion.

Another potential threat could be raised from the number of participants in the card sorting and focus groups. The categorization of AI incidents and the decision tree were validated through card sorting and focus group exercises, but these methods involved a small number of participants. Findings from this validation may not be generalizable.

\textbf{Dependability.} Dependability concerns the repeatability of the research findings in similar contexts. There exists a threat to the dependability of our study owing to it's largely qualitative nature. Our categorization, decision tree development, and subsequent analysis are driven mostly by interpretation, which may vary among different researchers. Additionally, the general dynamics of AI and D\&I issues are rapidly evolving, which may invalidate our decision tree and categorization over time. Therefore, it is essential that we persistently update and enrich our repository with new AI incidents. Additionally, we are planning to create the second version of our D\&I-related AI incidents database by merging data from other AI incident databases such as OECD.

\textbf{Transferability.} Transferability refers to the extent to which the findings can be applied in different contexts. The D\&I issues are not universally defined and may vary across different cultures, communities, and regions. Therefore, the AI incidents identified as related or unrelated to D\&I issues in our study may not be perceived the same way in other contexts. However, we proposed the methods to identify D\&I issues in AI incidents, that mitigate the threat. The methods can be applied in any AI incidents from different cultural settings.

%% file: Sections/7_Conclusions.tex
\section{Conclusions and Future Work}
\label{sec:conclusions}

With the aim to develop more inclusive, unbiased, and trustworthy AI systems, this study is a critical first step in identifying D\&I-related incidents in AI and understanding the extent to which D\&I issues exist within AI systems. We developed a decision tree as a preliminary framework for identifying and categorizing AI incidents in terms of their relation to D\&I issues. We also proposed and populated a public repository on D\&I-related AI incidents. The AI systems were found to have a significant association with D\&I issues, with 34.3\% and 46.45\% of the analyzed incidents being related to D\&I from AIID and AIAAIC databases respectively. This emphasizes the need for careful attention to D\&I during AI system design, development, and deployment. Moreover, the lack of information in some cases also complicates the task of categorizing AI incidents, making it difficult to draw definitive conclusions regarding their D\&I implications. Hence, a more informative and comprehensive representation of AI incidents is required. Despite the comprehensive consideration of 16 diversity attributes, findings indicated a concerned prominence of racial, gender, and age discrimination in AI incidents. It underlines the urgency to address `race', `gender', and `age' biases in AI system development, while not undermining the criticality of other attributes.

This study also provides a roadmap for future research in D\&I within AI. Given the dynamic nature of AI, there is a continuous need to populate and revise the repository with new incidents for timely research investigations. Additionally, we must analyze all the 575 existing AIAAIC incidents in future and include the D\&I-related AI incidents in our repository. Future studies should also try to understand why some AI incidents needed ``more information''. This could help figure out exactly what challenges or complications are making it difficult to clearly classify these incidents. Additionally, further research is essential to develop comprehensive guidelines, and concrete strategies to prevent the occurrence of D\&I-related AI incidents. Therefore, comprehensive guidelines are also necessary to embed D\&I principles into AI's design, development, and deployment.